\documentclass{elsart}
\usepackage{epsfig,amsmath,amssymb,color}

\begin{document}

\begin{frontmatter}

\title{Slow-light enhanced light-matter interactions with applications to gas sensing}

\author{K.~H. Jensen}

\address{Center for Fluid Dynamics, Department of Micro and Nanotechnology, Technical
University of Denmark, DTU Nanotech, Building 345 East, DK-2800
Kongens Lyngby, Denmark}

\author{M.~N. Alam}

\address{Department of Chemical and Biochemical Engineering, Technical
University of Denmark, DTU KT Building 227, DK-2800 Kongens Lyngby,
Denmark}

\author{B. Scherer and A. Lambrecht}

\address{Fraunhofer Institute for Physical Measurement Techniques, Heidenhofstrasse 8, D-79110 Freiburg, Germany}

\author{N.~A. Mortensen\corauthref{cor1}}

\ead{asger@mailaps.org}

\address{Department of Photonics Engineering, Technical
University of Denmark, DTU Fotonik, Building 345 West, DK-2800
Kongens Lyngby, Denmark}

\corauth[cor1]{Corresponding author}

\begin{abstract}
Optical gas detection in microsystems is limited by the short micron
scale optical path length available. Recently, the concept of
slow-light enhanced absorption has been proposed as a route to
compensate for the short path length in miniaturized absorption
cells. We extend the previous perturbation theory to the case of a
Bragg stack infiltrated by a spectrally strongly dispersive gas with
a narrow and distinct absorption peak. We show that considerable
signal enhancement is possible. As an example, we consider a Bragg
stack consisting of PMMA infiltrated by O$_2$. Here, the required
optical path length for visible to near-infrared detection ($\sim
760$~nm) can be reduced by at least a factor of $10^2$, making a
path length of 1~mm feasible. By using this technique, optical gas
detection can potentially be made possible in microsystems.
\end{abstract}

\end{frontmatter}

%------------------------------------------------------------------
\section{Introduction}
\begin{figure}
\begin{center}
\epsfig{file=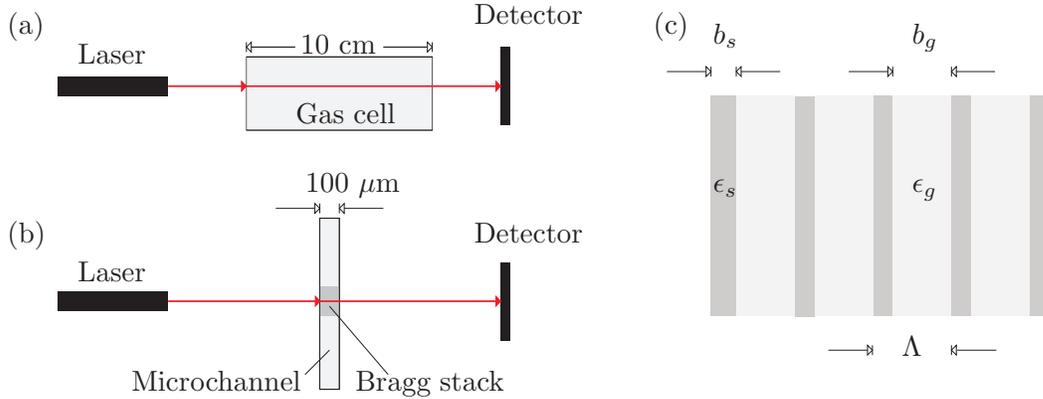, width=1.0\textwidth,clip,angle=0}
\end{center}
\caption{\label{fig:setup} (a) Schematic of a typical macroscopic
gas detection setup with a light-matter interaction path length of
$10$~cm. (b) Schematic of a microscopic gas detection setup with a
light-matter interaction path length of $100\,{\rm \mu m}$ made
possible by the presence of a Bragg stack inside the microchannel.
(c) Liquid/gas infiltrated Bragg stack composed of alternating solid
and gas layers of thickness $b_s$ and $b_g$, respectively.}
\end{figure}

%------------------------------------------------------------------
The integration of optics and microfluidics on lab-on-a-chip
microsystems has recently been the topic of much
research~\cite{Erickson:2008,Monat:2007,Psaltis:2006}, partly
motivated by its diverse applications in chemical and biochemical
analysis~\cite{Janasek:2006}. The miniaturization of chemical
analysis systems, however, presents many optical challenges since
light matter interactions suffer from the reduced optical path
length $L$ in lab-on-a-chip systems compared to their macroscopic
counterparts as illustrated in Fig.~\ref{fig:setup}. Mogensen
\emph{et al.}~\cite{Mogensen:2003} demonstrated that for
Beer--Lambert absorption measurements in a lab-on-a-chip system, a
typical size reduction by two orders of magnitude severely reduces
the optical sensitivity in an inversely proportional manner. This
drawback is even more pronounced for gas detection in microsystems
as most gases have a very weak absorption line. Thus, microsystems
may not seem the obvious solution for gas sensing and detection. On
the other hand, the major reason for being interested in pursuing
microsystem opportunities is of course that many applications call
for compact gas sensors because of limited sensor space. On the more
practical level, we emphasize that lab-on-a-chip systems can both
address very minute volumes (no flow), but also allows for
measurements of larger volumes flowing through the microchannels
where detection takes place. In that way, spatial variations in the
gas properties can potentially be mapped onto a time axis and
monitored through time-traces of the optical signal. This space-time
mapping is a general approach (employed for a variety of other
measurements) which is only possible in microchannels supporting
laminar flow. For the optical sensitivity, it has been proposed,
that by introducing a porous and strongly dispersive material, such
as a photonic crystal, into the lab-on-a-chip microsystems one could
potentially solve these problems through slow-light enhanced
light-matter
interactions~\cite{Mortensen:2007,Pedersen:2007a,Pedersen:2007,Mortensen:2008}.
First experimental evidence of this effect for gas sensors in the
mid-infrared range was reported by Lambrecht \emph{et
al.}~\cite{Lambrecht:2007}. Here, we extend the perturbative theory
in
Refs.~\cite{Mortensen:2007,Pedersen:2007a,Pedersen:2007,Mortensen:2008}
to the case of a dispersive Bragg stack infiltrated by a spectrally
strongly dispersive gas with a narrow and distinct absorption peak.
We emphasize the example of oxygen monitoring and sensing based on
the two distinctive bands near the visible to near-infrared light
range of 760~nm referred to as the O$_2$ A band~\cite{Scherer:2008}.
Spectral data are available from the HITRAN
database~\cite{Rothman:2005}. Though, the A band has a very weak
absorption feature, it offers the potential to establish an optical
\emph{in situ} O$_2$ detection for application in many fields
including combustion processes~\cite{Schlosser:2003} and fire
research~\cite{Laakso:2005}.
\section{Beer--Lambert absorption}
In gas detection experiments, one typically uses Beer--Lambert
absorption to determine the concentration of the substance of
interest. As the light traverses the test chamber the light
intensity is exponentially attenuated according to the relation
\begin{equation}\label{eq:Iexp}
I=I_0\exp (-\alpha L),
\end{equation}
where $\alpha$ is the absorption coefficient, typically a linear
function of the concentration ${\mathcal C}$ of absorbing molecules
(not to be confused with $c$; the speed of light in vacuum to be
introduced at a later stage). Any reduction in the optical path
length $L$ will penalize the sensitivity of the optical measurement
significantly. However, by using a porous photonic crystal, it has
been suggested that one may effectively enhance $\alpha$ by slowing
down the light via a photonic crystal thereby overcoming the short
path length obstacle in microsystems~\cite{Mortensen:2007}. To
quantify the enhancement, we shall use the ratio
\begin{equation}\label{eq:gamma1}
\gamma=\frac{\Delta\alpha_{\rm pc}}{\alpha_g},\quad
\Delta\alpha_{\rm pc}=\alpha_{\rm pc}-\lim_{\alpha_g\rightarrow
0}\alpha_{\rm pc},
\end{equation}
where $\alpha_{\rm pc}$ is the absorption coefficient inside the
photonic crystal and $\alpha_g$ the corresponding homogeneous space
absorption coefficient associated with the gas itself. The
subscripts are introduced in order to carefully distinguish between
absorption taking place in the photonic crystal material and the
gas, respectively. In this work we consider the particular simple
one-dimensional example of a Bragg stack and extend previous
work~\cite{Pedersen:2008a} by using a more realistic model for the
gas absorption profile.
\section{The Bragg stack}
The periodic Bragg stack is illustrated in panel~(c) of
Fig.~\ref{fig:setup} and is composed of alternating layers of gas
and solid of thickness $b_g$ and $b_s$, respectively. The lattice
constant is $\Lambda=b_g+b_s$ and for the dielectric functions we
allow for complex-valued bulk parameters, i.e.
$\epsilon_g=\epsilon_g'+i\epsilon_g''$ and
$\epsilon_s=\epsilon_s'+i\epsilon_s''$. For later convenience we
rewrite the dielectric functions in terms of the bulk refractive
index $n$ and the damping coefficient $\alpha$,
\begin{subequations}
\begin{equation}
\epsilon_g=n_g^2+in_g\alpha_g/k,
\end{equation}
\begin{equation}
\epsilon_s=n_s^2+in_s\alpha_s/k,
\end{equation}
\end{subequations}
where $k=2\pi/\lambda=\omega/c$ is the free-space wave vector with
$\lambda$ being the corresponding free-space wavelength and $\omega$
the angular frequency. The dispersion relation for the Bragg stack
is governed by (see Ref.~\cite{Pedersen:2008a} and references
therein)
\begin{subequations}
\begin{equation}
 \cos \left(\kappa \Lambda\right) = F(k),
 \label{eq:disprel}
\end{equation}
where $\kappa$ is the Bloch wave vector and
\begin{equation}
F(k)=\cos\left(\sqrt{\epsilon_g}kb_g \right)\cos\left(
\sqrt{\epsilon_s}kb_s\right)
-\frac{\epsilon_g+\epsilon_s}{2\sqrt{\epsilon_g \epsilon_s}}
\sin\left( \sqrt{\epsilon_g}kb_g\right) \sin\left(
\sqrt{\epsilon_s}kb_s \right). \label{eq:F}
\end{equation}
\end{subequations}
In the following we explicitly write the Bloch wave vector as
$\kappa\equiv\kappa'+i\kappa''$, where $\kappa'$ and $\kappa''$ are
real and imaginary parts, respectively. The imaginary part causes an
exponential damping of the Bloch states with a corresponding
attenuation coefficient $\alpha_{\rm pc}=2\kappa''$. Likewise, in
the homogeneous case we have
$\alpha_g=k\epsilon_g''/\sqrt{\epsilon_g'}$ and in this way
Eq.~(\ref{eq:gamma1}) now becomes
\begin{equation}\label{eq:gamma2}
\gamma=\frac{2\sqrt{\epsilon_g'}}{\epsilon_g''}\times\frac{\kappa''(\epsilon_s'',\epsilon_g'')-\kappa''(\epsilon_s'',\epsilon_g''\rightarrow
0)}{k}.
\end{equation}
If $\gamma$ is greater than unity, it means that we have enhanced
absorption due to the geometry of the Bragg stack. The absorption
enhancement comes about because the group velocity of the light
$\partial \omega/\partial \kappa'$ is reduced by the geometry of the
Bragg stack~\cite{Mortensen:2007}. In Eq.~(\ref{eq:gamma2}), the
slow light phenomenon manifests itself through an increasing
imaginary part of the Bloch vector $\kappa''$ as the band edge is
approached. In deriving Eq.~(\ref{eq:gamma2}), we have separated the
two different contributions to the absorption from the gas and the
solid, thus implicitly assuming a small change in attenuation caused
by the presence of the gas. However, we emphasize that the effect of
material dispersion is included in the dispersion relation in an
exact analytical and non-perturbative way and thus our results
fulfill the Kramers--Kronig relation by taking the full interplay
between waveguide and material dispersion into account.
\section{Examples}
In the following we illustrate the prospects for slow-light enhanced
absorption by means of examples.
\subsection{Non-absorbing materials}
\begin{figure}
\begin{center}
\epsfig{file=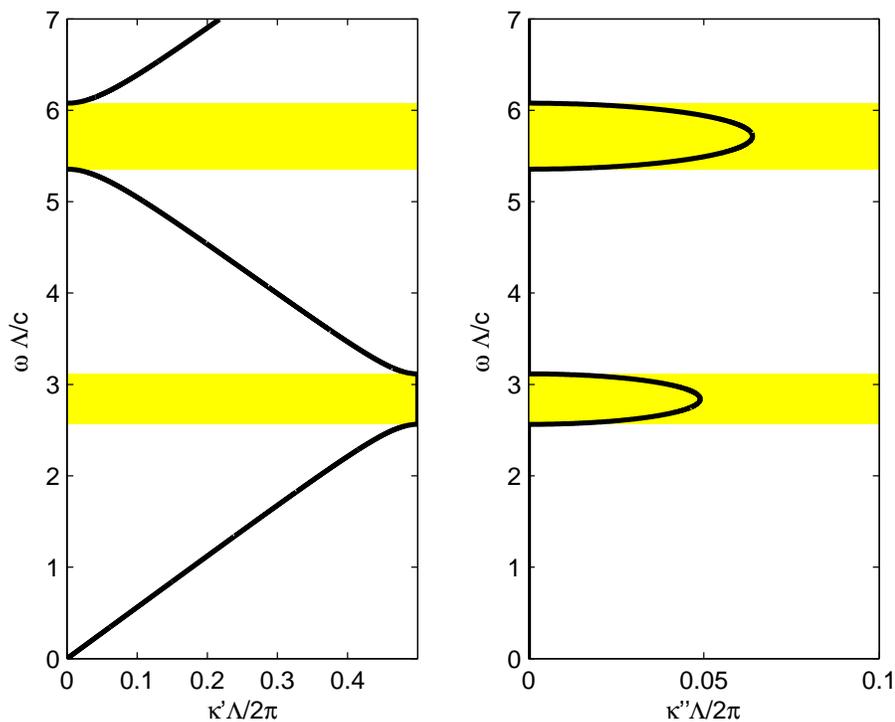, width=\textwidth,clip,angle=0}
\end{center}
\caption{\label{fig:bandiag1}Band diagram for a Bragg stack with
$\epsilon_g=1.0^2$, $\epsilon_s=1.5^2$, $b_g=0.8\Lambda$, and
$b_s=0.2\Lambda$, corresponding to a polymer infiltrated by gas.}
\end{figure}

First, we consider the problem of two non-absorbing materials, say a
gas and a polymer with $\epsilon_g=1.0^2$ and $\epsilon_s=1.5^2$.
The corresponding band diagram, obtained from a numerical solution
of Eq.~(\ref{eq:disprel}), is shown in Fig.~\ref{fig:bandiag1}. Two
band gaps, indicated by yellow shading, are visible near
$\omega\Lambda/c=2.8$ and $5.5$. As can be seen, the group velocity
approaches zero near the band gap edges while the imaginary part of
$\kappa$ attains a finite value only inside the band gaps where
propagation of electromagnetic radiation is prohibited due to the
exponential damping.
\subsection{Frequency-independent weakly absorbing materials}
\begin{figure}
\begin{center}
\epsfig{file=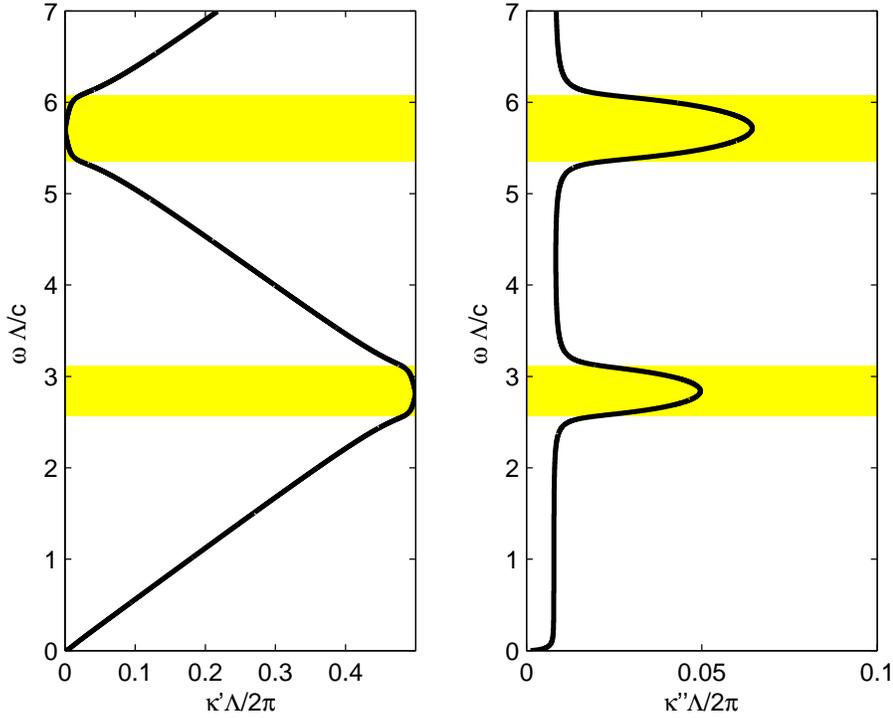, width=\columnwidth,clip,angle=0}
\end{center}
\caption{\label{fig:bandiag3} Band diagram for the same problem as
in Fig.~\ref{fig:bandiag1}, but with
$\alpha_g\Lambda=\alpha_s\Lambda=0.1$, corresponding to a finite
bulk damping in the gas and the polymeer, respectively.}
\end{figure}

Next, we consider two weakly absorbing materials, say again the same
gas and polymer, but now with imaginary contributions to the
dielectric functions corresponding to damping parameters
$\alpha_g\Lambda=\alpha_s\Lambda=0.1$. As seen from the
corresponding band diagram in Fig.~\ref{fig:bandiag3}, the main
cause is to smear out the spectrally pronounced dispersion features.
First, it increases $\kappa''$ from zero to a finite value outside
the band gaps and second it smears out the band gap, setting a lower
limit on the magnitude of the group velocity.
\subsection{Strongly frequency-dependent absorbing gases}
Having treated the simple case of spectrally uniform absorption, we
shall now consider a slightly more realistic model for the
frequency-dependent absorption in gasses. Due to causality and the
Kramers--Kronig relation, the presence of frequency-dependent
absorption is inevitably accompanied by material dispersion in the
gas which will play in concert with the waveguide dispersion
contributed by the geometry of the Bragg stack. To this end, we will
now consider a Lorentzian absorption line of the
form~\cite{Griffiths:99}
\begin{subequations}
\begin{equation}
 \epsilon_g''(\omega)=A\frac{\sigma\omega}{
 \left(\omega^2-\omega_0^2\right)^2+\sigma^2\omega^2},
\label{eq:abs1}
\end{equation}
where the corresponding real part of the dielectric function follows
from the Kramers--Kronig relation and is given by
\begin{equation}
 \epsilon_g'(\omega)=1+A\frac{\omega_0^2-\omega^2}{
 \left(\omega^2-\omega_0^2\right)^2+\sigma^2\omega^2}.
\label{eq:abs2}
\end{equation}
\end{subequations}
In this model, the strength $A$ scales with the concentration of
molecules while $\sigma$ quantifies the spectral line width around
the resonance frequency $\omega_0$.
\begin{figure}[t!]
\begin{center}
\epsfig{file=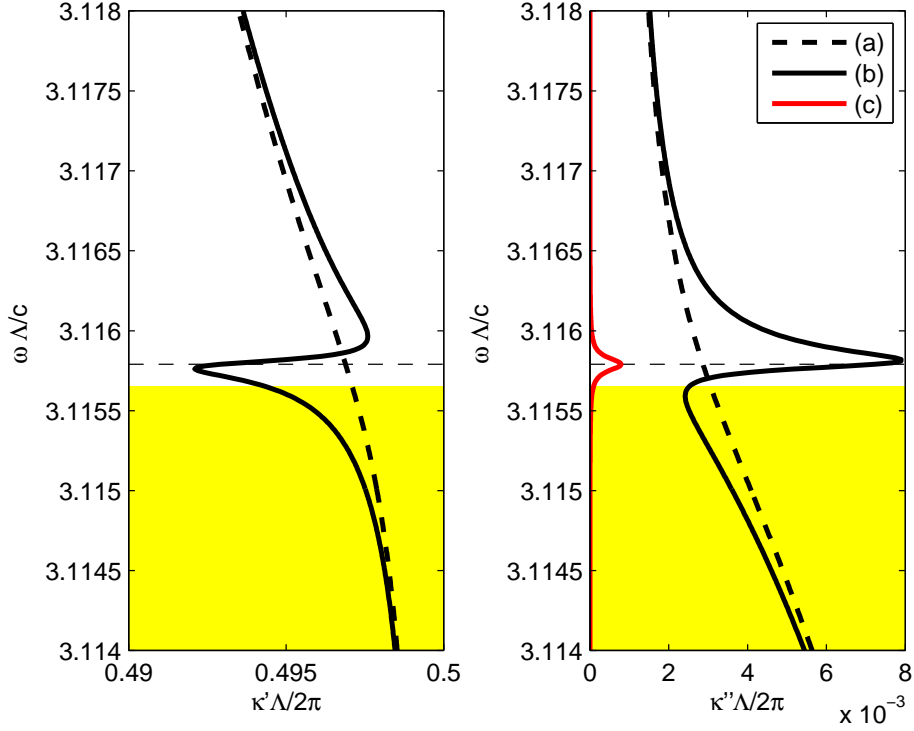, width=\columnwidth,clip,angle=0}
\end{center}
\caption{\label{fig:banddiag3} Band diagram for a Bragg stack with
$\alpha_s\Lambda=0.1$, $n_s=1.5$, $b_g=0.8\Lambda$, $b_s=0.2\Lambda$
and $\epsilon_g$ given by equations \ref{eq:abs1}-\ref{eq:abs2} with
$\omega_0\Lambda/c=3.1158,\;\sigma\Lambda/c=10^{-4}$ and
$Ac^2/\Lambda^2=10^{-6}$. The black solid line (b) shows the results
for the gas-infiltrated Bragg stack, while the corresponding dashed
line (a) is in the absence of gas infiltration. In the $\kappa''$
plot, the red solid line (c) shows the absorption in the absence of
the Bragg stack. Only the spectral region just above the band gap is
emphasized in the plots.}
\end{figure}
As an example the band diagram for $\alpha_s\Lambda=0.1$,
$\omega_0\Lambda/c=3.1158$, $\sigma\Lambda/c=10^{-4}$, and $A
c^2/\Lambda^2=10^{-6}$ is shown in Fig.~\ref{fig:banddiag3}. Note
the zoom-in on the first band gap and the different scale compared
to Figs.~\ref{fig:bandiag1} and \ref{fig:bandiag3}. The horizontal
dashed black line indicates the above-band-edge position of the
absorption line while the yellow shading indicates the band gap for
the absorption-free structure. The red line shows bulk-absorption of
the gas itself, thus indicating an absorption enhancement of the
order $\gamma\simeq 5.5$. We emphasize that this value depends
strongly on $\alpha_g$, $\alpha_s$, $b_g$, and $b_s$ and as we shall
see in the next section, even stronger absorption enhancement is
possible by carefully tuning the Bragg stack parameters to fit the
absorption line of interest.
\section{Application to oxygen detection}
Oxygen molecules, O$_2$, exhibit a distinctive absorption band in
the visible to near-infrared light range at 760~nm, which is called
the Oxygen A band or the atmospheric transition of oxygen. These
transitions are electric dipole forbidden, so only magnetic
transitions are allowed, that require a change in the rotational
quantum number of $dN=\pm 1$. Therefore the A-band splits in to the
R-branch ($dN=-1$), starting at 759.8~nm with a bandwidth of 2.2~nm
and the P-branch ($dN=+1$) starting at 762.3~nm with a bandwidth of
5.7~nm~\cite{Brown:2000,Scherer:2008}. Though, the A band has a very
weak absorption feature, it offers the potential to establish an
optical \emph{in situ} O$_2$ detection for a variety of
applications, ranging from combustion
processes~\cite{Schlosser:2003} and fire research~\cite{Laakso:2005}
to biomedical and environmental systems~\cite{Raab:1981}. For this
purpose different spectroscopic methods are used to enlarge the
sensitivity of absorption measurements. Folding the laser beam
several times in a multi reflection cell or using the long photon
life times of high finesse cavities (cavity ring-down spectroscopy)
offer an enhancement in sensitivity in respect to an enlargement of
the absorption path length. In contrast, the method of slow light
enhancement is related to an increased light matter interaction
(though it may be phenomenologically explained as an enlarged
effective path length), leading to higher absorption cross sections.
This effect offers the potential for micrometer scaled probe volumes
which is an important feature for e.g. lab on a chip applications.
Typically, the absorption of oxygen in ambient air at the 760~nm
line is of the order 2.6\% with an optical path length of around
1~m~\cite{Schlosser:2003}. At this wavelength, the O$_2$ absorption
spectrum is minimally sensitive towards temperature changes and it
is fairly influenced by the pressure
variation~\cite{Brown:2000,Scherer:2008,Schlosser:2003}. The lowest
oxygen concentration that can be measured is about 20-50~ppm which
corresponds to transmission change of about
0.0005~\cite{Scherer:2008}. The smallest experimentally detected
absorption is about 3 times larger than the shot noise limit and is
about $10^{-6}$. This corresponds to a concentration of about 8~ppm
at 1~m path length~\cite{Vogel:2001,Werle:1989}.

\begin{figure}
\begin{center}
\epsfig{file=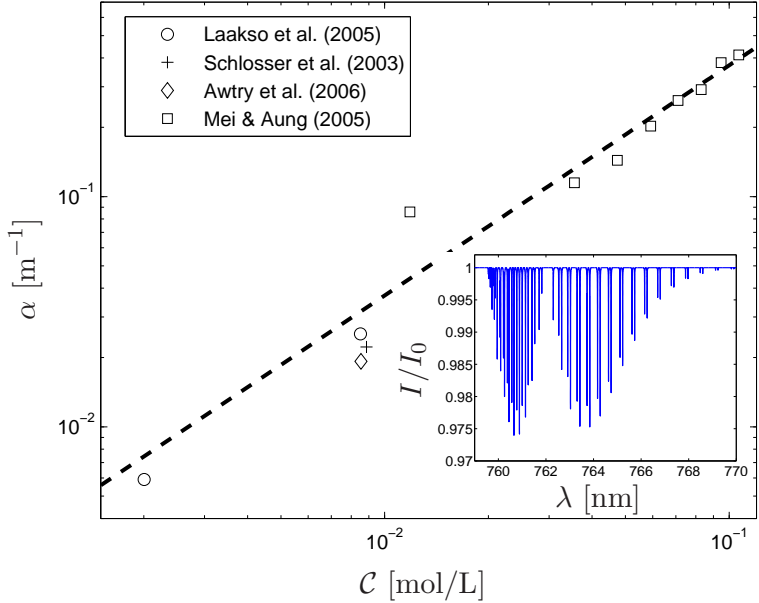, width=0.8\columnwidth,clip,angle=0}
\end{center}
\caption{\label{fig:o2alpha}O$_2$ absorption coefficient $\alpha$ as
a function of the O$_2$ concentration ${\mathcal C}$ in the radiated
volume. The inset shows the A band based on HITRAN
data~\cite{Rothman:2005} for a path length of 1~m.}
\end{figure}

By extracting the spectroscopic data from the published
works~\cite{Schlosser:2003,Laakso:2005,Robichaud:2008,Awtry:2006,Mei:05},
we correlated the O$_2$ absorption coefficient, $\alpha$ [see
Eq.~(\ref{eq:Iexp})] as a function of O$_2$ concentration ${\mathcal
C}$ in the radiated volume. To lowest order in the gas concentration
we expect that $\alpha({\mathcal C})=\eta {\mathcal C}$ and the
linearity of the correlation obtained is shown in
Fig.~\ref{fig:o2alpha}. The slope of the relatively linear line
shows that it is possible to obtain a fairly good measurement of
O$_2$ molecule through the excitation of light at $\lambda \sim
760\,{\rm nm}$. From a linear fit we find that $\eta\sim 3.7\,({\rm
mol/L})^{-1} {\rm m}^{-1}$. We note that the different absorption
lines in the A-band have different line strengths which may cause
the scatter of data in Fig.~\ref{fig:o2alpha}.
The absorption in PMMA, at frequencies around the O$_2$ A band of
oxygen, is fairly constant around $10/$m, i.e. of the order of
$2.5\times 10^2$ times stronger than the absorption in
O$_2$~\cite{Khanarian:2001}.

For the absorption coefficient in O$_2$, we note that the $2.6\%$
damping over one meter in the framework of Eq.~(\ref{eq:Iexp})
corresponds to $I/I_0=1-0.026=\exp(-0.20942\times
\alpha_{\text{O}_2}\times 1~m)$ where we have explicitly accounted
for the volume fraction of oxygen being $20.942\%$ in ambient air.
This translates into $\alpha_{\text{O}_2}=0.126\,{\rm m}^{-1}$ and
for PMMA we have $\alpha_{\text{PMMA}}=10\,{\rm m}^{-1}$ while the
corresponding dielectric functions are $\epsilon'_{\text{O}_2}=1^2$
and $\epsilon'_{\text{PMMA}}=1.5^2$. Choosing $\Lambda=1\,{\rm \mu
m}$ we find that the oxygen A band will be located right next to a
band gap edge if we choose $b_g=0.6262\Lambda$, resembling the
situation shown in Fig.~\ref{fig:banddiag3}. Assuming a Lorentzian
absorption profile and using the method discussed above, we find
that $\gamma_{\text{O}_2}\simeq 235$. We emphasize that the
Lorentzian linewidth correspond to the complete A band and thus
resembles a group of different and even sharper absorption lines,
see inset of Fig.~\ref{fig:o2alpha}, for which the enhance factor
varies.

In a typical Beer--Lambert experiments measuring the O$_2$
concentration in the atmosphere, a path length of around 100 cm
gives a 2.6\% attenuation of the transmitted signal. The enhanced
absorption in the Bragg stack discussed above implies that one could
scale the path length down by a factor of
$235\frac{b_g}{\Lambda}\simeq 145$ to about 7~mm while retaining the
same absorption signal. We emphasize that this sensitivity
enhancement factor (effective path length/ real length of the
absorber) is even higher than that of a multi reflection cell, which
is limited by reflection losses to about 100! Very high enhancement
factors of 5 orders-of-magnitude or more compared to a single pass
setup may be achieved using cavity-enhanced or cavity ringdown laser
spectroscopy. However, such macrooptical setups can hardly be
condensed into a microoptical device, and are generally delicate
instruments mainly operated in laboratory environments. Using pairs
of highly reflecting mirrors with reflectivities exceeding 99.999\%,
these setups are very sensitive to contamination e.g. by dust. In
contrast the slow-light concept can be considered as a manifold of
distributed mirrors. Hopefully, contamination shall not be as
detrimental as with the macroscopic cavity approaches. In terms of
bandwidth, the slow-light phenomenon shares physics with the various
cavity-based approaches: strong enhancement of the absorption is
obtained at the price of a narrow band-width!

\section{Conclusion}
Slow-light phenomena in photonic crystals has recently been proposed
as a mechanism to allow for reduced path length in absorption
measurements in microsystems. In this paper we have considered the
effect an absorption line in a gas infiltrating a uniformly
absorbing Bragg stack. We have shown that enhanced absorption is
possible and as an example we have considered a Bragg stack
consisting of PMMA infiltrated by O$_2$. In that case we have found
the absorption enhancement to exceed a factor of $10^2$ potentially
allowing for O$_2$ detection with sub millimeter optical path
lengths. For future directions we emphasize studies of the influence
of the interfaces (scattering loss), structural imperfections, and
beam divergence in more detail and another natural extension would
be planar two-dimensional photonic crystals membranes.

\section*{Acknowledgement}
We acknowledge Martin Heller and Henrik Bruus for useful discussions
and technical assistance. Publication of this work is financially
supported by the \emph{Danish Research Council for Technology and
Production Sciences} (grant no: 274-07-0080). KHJ acknowledges
financial support by the \emph{Danish National Research Foundation}
(grant no: 74) and MNA is financially supported by \emph{Universiti
Teknologi Malaysia} and the \emph{Malaysia Ministry of High
Education}.

\newpage
%\bibliographystyle{elsart-num}
%\bibliography{Bibtex_biography}

\end{document}